\documentclass[prd,preprint,superscriptaddress,amsmath,amssymb,nofootinbib]{revtex4}
\usepackage{graphicx}
\usepackage{dcolumn}
\usepackage{bm}
\usepackage{amssymb}
\usepackage{amsmath}
\usepackage{epsfig}    
\usepackage{color}
\usepackage{slashed}
\usepackage{hhline}

\def\be{\begin{equation}}
\def\ee{\end{equation}}
\newcommand{\bea}{\begin{eqnarray}}
\newcommand{\eea}{\end{eqnarray}}
\newcommand{\nn}{\nonumber}

\begin{document}

{\begin{flushright}{KIAS-P20041,  APCTP Pre2020 - 016}
\end{flushright}}

\title{A two-loop induced neutrino mass model, dark matter, and LFV processes $\ell_i \to \ell_j \gamma$, and $\mu e \to e e$  in a hidden local $U(1)$ symmetry} 

\author{Takaaki Nomura}
\email{nomura@kias.re.kr}
\affiliation{School of Physics, KIAS, Seoul 02455, Korea}

\author{Hiroshi Okada}
\email{hiroshi.okada@apctp.org}
\affiliation{Asia Pacific Center for Theoretical Physics (APCTP) - Headquarters San 31, Hyoja-dong,
Nam-gu, Pohang 790-784, Korea}
\affiliation{Department of Physics, Pohang University of Science and Technology, Pohang 37673, Republic of Korea}

\author{Yuichi Uesaka}
\email{uesaka@ip.kyusan-u.ac.jp}
\affiliation{Faculty of Science and Engineering, Kyushu Sangyo University, 2-3-1 Matsukadai, Higashi-ku, Fukuoka 813-8503, Japan}

\date{\today}

\begin{abstract}
We discuss a  model based on a hidden $U(1)_X$ gauge symmetry in which neutrino mass is induced at two-loop level by effects of 
interactions among particles in hidden sector and the Standard Model leptons.
Since neutrino mass is suppressed by two-loop, its associate Yukawa couplings can be sizable and it would affect lepton flavor phenomenology.
We analyze neutrino mass matrix, lepton flavor violating processes, electron/muon $g-2$ and dark matter annihilation cross section
 which are induced via interactions among Standard Model leptons and particles in $U(1)_X$ hidden sector, and their interactions can be sizable in our scenario.
Performing numerical analysis, we show expected ratios for these processes using allowed parameters which can fit the neutrino data and satisfy flavor constraints.
\end{abstract}
\maketitle
\newpage

\section{Introduction}
There are some issues requiring physics beyond the standard model (SM) such as a mechanism of generating non-zero neutrino masses and existence of dark matter(DM).
One of the most attractive scenarios connecting these issues is a radiative neutrino mass generation in which neutrino mass is realized at loop level~\cite{Ma:2006km}. 
In such a case a diagram of neutrino mass generation is often induced by particles in hidden sector including DM candidate.
For describing a hidden sector, an introduction of hidden $U(1)$ symmetry is one of the most attractive possibilities to forbid tree level neutrino mass and stabilize DM candidate~\cite{Dey:2019cts, Nomura:2018kdi, Cai:2018upp, Nomura:2018ibs, Nomura:2017wxf, Ko:2017uyb, Ko:2016uft, Ko:2016wce, Ko:2016ala, Ko:2014loa, Ko:2014eqa,Ma:2013yga,Yu:2016lof,Ko:2016sxg, Nomura:2020azp,Nomura:2020twp,Kim:2020aua}.

In a radiative neutrino mass generation model,  
tiny neutrino mass can be realized naturally due to loop suppression factor and we would have sizable Yukawa interactions between hidden particles and SM leptons.
In particular, Yukawa couplings tend to be larger when neutrino masses are generated at higher loop like two-loop level.
Interestingly,  when these couplings are sizable we can have rich flavor phenomenology such as lepton flavor violating (LFV) processes  $\ell_i \to \ell_j \gamma$, $\ell_i \to \ell_j \ell_k \bar \ell_l$ and $\mu e \to e e$.
In addition we could obtain sizable anomalous dipole magnetic moment of electron and muon (electron/muon $g-2$).

In this paper, we construct a model with hidden sector based on a local $U(1)_X$ symmetry in which neutrino mass is generated at two-loop level.
In addition scalar singlet is introduced as a DM candidate whose stability is guaranteed by $Z_2$ symmetry as a remnant of $U(1)_X$.
Neutrino masses are suppressed by two-loop factor and their Yukawa couplings for neutrino mass generation can be sizable.
We analyze neutrino mass matrix, LFV processes, electron/muon $g-2$ and DM annihilation cross section which are induced via interactions among SM leptons and particles in $U(1)_X$ hidden sector.
Carrying out numerical analysis we search for allowed parameter region and estimate expected ratios for various LFV processes and electron/muon $g-2$.

This paper is organized as follows.
In Sec.~II, we show our model and analyze 
neutrino mass generation mechanism at two-loop level,  LFVs  and  electron/muon $g-2$.
In Sec.~III, we perform numerical analysis searching for allowed parameter sets and 
estimate ratios of LFV processes, electron/muon $g-2$ and DM annihilation cross section with these parameters.
In Sec.~IV, we give summary of our results and conclusion.

\section{Model}

\begin{table}[t!]
\begin{tabular}{|c||c|c|c||c|c|c|c|c|c|}\hline\hline  
& ~$L_L$~& ~$e_R$~  & ~$E$~& ~$H$~& ~$\eta$~ & ~$s^+$~& ~$k^{++}$~ & ~$\varphi$~ & ~$\chi$~ \\\hline
$SU(2)_L$ & $\bm{2}$  & $\bm{1}$  & $\bm{1}$  & $\bm{2}$  & $\bm{2}$  & $\bm{1}$   & $\bm{1}$   & $\bm{1}$& $\bm{1}$    \\\hline 
$U(1)_Y$   & $-\frac12$ & $-1$ & $-1$ & $\frac12$  & $\frac12$  & $1$ & $2$ & $0$ & $0$  \\\hline
$U(1)_{X}$   & $0$ & $0$ & $Q_X$   & $0$  & $-Q_X$  & $-Q_X$  & $-2Q_X$ & $2Q_X$  & $Q_X$\\\hline
\end{tabular}
\caption{ 
Charge assignments to fields in the model under $SU(2)_L\times U(1)_Y\times U(1)_X$ where we omitted quark sector since it is the same as the SM one.  }
\label{tab:1}
\end{table}

In this section, we extend the SM into a hidden $U(1)_X$ gauged symmetry.
At first, we introduce three families of singly-charged exotic fermions  $E$ with $U(1)_X$ charge of $Q_X$.
Then, we introduce an isospin inert doublet boson $\eta = (\eta^+, \eta^0)^T$ with $-Q_X$ charge, an isospin singlet  singly-charged one $s^+$, an isospin singlet  doubly-charged one $k^{++}$, an isospin singlet neutral one $\varphi$ and another isospin singlet scalar $\chi$ each of which has $-Q_X$, $-2Q_X$, $2Q_X$ and $Q_X$ hidden charges.
$\varphi$ has nonzero vacuum expectation value (VEV) which is denoted by $v'/\sqrt2$, where the SM Higgs is symbolized by $H$ whose VEV is written by $v/\sqrt2$.
The SM singlet scalar $\varphi$ breaks the hidden $U(1)$ gauge symmetry spontaneously by developing its VEV.
Notice that $U(1)_X$ breaks into $Z_2$ symmetry in which $E$, $\eta$ and $\chi$ are odd and the other fields are even; 
this remaining symmetry guarantees the stability of our dark matter candidate which is chosen to be $\chi$ in our scenario.
The charge assignments of our fields is summarized in Table~\ref{tab:1} where quark sector is abbreviated, since they are the same as the SM charge assignment.
Under the symmetries in a renormalized theory, the relevant Lagrangian is given by 
\begin{align}
& -{\cal L_M}  =   M_{E} \bar E_L E_R + {\rm h.c.},  \label{Eq:Mass} \\
& -{\cal L_\ell}
= y_{\ell} \bar L_L H e_R  +  f  \bar L_L\eta  E_R  +  g_R k^{++} \bar E_R^C E_R +  g_L k^{++} \bar E_L^C E_L 
+ h \bar E_L e_R \chi + {\rm h.c.}, \label{Eq:yuk} 
\end{align}
where we neglect the indices of families, and $y_\ell$ is supposed to be a diagonal matrix without loss of generality due to the redefinitions of the fermions.
The scalar potential is also given by
\begin{align}
V = & \mu_H^2 H^\dagger H +\mu_\eta^2 \eta^\dagger\eta + \mu_s^2 |s^+|^2+ \mu_k^2 |k^{++}|^2 + \mu_\varphi^2 |\varphi|^2 \\
&+\mu[ (H^Ti\sigma_2 \eta) s^-+ {\rm h.c.}]+ \frac12 \mu_{kss} [k^{++} s^- s^-  + {\rm h.c.}] + \lambda_{X} ( H^\dag \eta \chi^* \varphi + h.c.) \nonumber \\
&  + \lambda_H (H^\dagger H)^2 + \lambda_\eta (\eta^\dagger \eta)^2 + \lambda_s (s^+ s^-)^2 + \lambda_k (k^{++} k^{--})^2
+ \lambda_{\varphi} (\varphi^* \varphi)^2  + \lambda_{H \eta} (H^\dagger H)(\eta^\dagger \eta) \nn \\
&+ \lambda'_{H \eta} (H^\dagger \eta)(\eta^\dagger H)
 + \lambda_{H s} (H^\dagger H)(s^+ s^-) + \lambda_{H k} (H^\dagger H)(k^{++} k^{--})
 + \lambda_{H \varphi} (H^\dagger H)(\varphi^* \varphi) \nn\\
 &+ \lambda_{\eta s}|\eta|^2 |s^+|^2+ \lambda_{\eta k}| \eta|^2 |k^{++}|^2+ \lambda_{\eta \varphi}|\eta|^2|\varphi|^2
+ \lambda_{sk}|s^+|^2|k^{++}|^2
+ \lambda_{s\varphi}|s^+|^2|\varphi|^2
+ \lambda_{k\varphi}|k^{++}|^2|\varphi|^2 , \nn
\end{align}
where $\sigma_2$ is the second Pauli matrix and we assume all couplings are real.

\subsection{Masses of extra bosons}

Here we discuss masses of extra scalar and gauge bosons.
The mass term for $(\eta^+, s^+)^T$ is obtained from the potential such as
\begin{equation}
\begin{pmatrix} \eta^+ \\ s^+ \end{pmatrix}^T \begin{pmatrix} m_\eta^2 & - \frac{\mu v}{\sqrt{2}} \\ - \frac{\mu v}{\sqrt{2}} & m_s^2 \end{pmatrix} \begin{pmatrix} \eta^- \\ s^- \end{pmatrix}, \
m_\eta^2 \equiv \mu^2_\eta + \frac{\lambda_{H \eta}}{2} v^2, \ m_s^2 \equiv \mu_s^2 + \frac{\lambda_{Hs}}{2} v^2.
\end{equation}
Thus $\eta^\pm$ and $s^\pm$ mix and  
mass eigenstates of singly charged-bosons are given by   
\begin{equation}
\begin{pmatrix} s^\pm \\ \eta^\pm \end{pmatrix} = \begin{pmatrix} \cos \alpha & -\sin \alpha \\  \sin \alpha & \cos \alpha \end{pmatrix} \begin{pmatrix} H^\pm_1 \\ H^\pm_2 \end{pmatrix}, \quad
\tan 2 \alpha = \frac{ - \sqrt{2} \mu v}{m_\eta^2 - m_s^2},
\label{eq:scalar-mass-fields}
\end{equation}
where $\alpha$ is the mixing angle that is taken to be free parameter in our numerical analysis.
Mass eigenvalues are also given by
\begin{align}
m_{H_1^\pm} & = \frac{1}{2} \left( m_s^2 + m_\eta^2 - \sqrt{(m_s^2 - m_\eta^2)^2 + 2 v^2 \mu^2} \right), \nonumber \\
m_{H_2^\pm} & = \frac{1}{2} \left( m_s^2 + m_\eta^2 + \sqrt{(m_s^2 - m_\eta^2)^2 + 2 v^2 \mu^2} \right),
\end{align}
where $m_{H_1^\pm} < m_{H_2^\pm}$ in our notation.

Note that we have a mixing between neutral component of $\eta$ and $\chi$ through $H^\dag \eta \chi^* \varphi$ term.
Here we write $\chi=\cos{\theta_s} H_1 -\sin{\theta_s} H_2$, and  $\eta^0=\sin{\theta_s} H_1 +\cos{\theta_s} H_2$; $H_{1,2}$ being the mass eigenstates and $m_{H_{1,2}}$ are their mass eigenvalues, respectively. 
{In our numerical analysis, we take $\sin \theta_s$ as free parameter and assume it is small as $\sin \theta_s < 0.01$ for simplicity.}

In addition, $Z_2$ even neutral scalar bosons from $H$ and $\varphi$ can mix since both fields develop VEVs.
In our analysis, such a mixing is taken to be small by assuming $\lambda_{H\varphi}$ to be tiny and 
$H$ is considered to be the SM-like Higgs.

After spontaneous symmetry breaking, we obtain massive $Z'$ boson whose mass is given by $m_{Z'} = 2 Q_X g_X v'$; $g_X$ is $U(1)_X$ gauge coupling.
In this paper we take $Z'$ mass is heavier than TeV scale and do not consider in our discussion of phenomenology.

\subsection{Neutrino mass generation}

 \begin{figure}[tb]
\includegraphics[width=80mm]{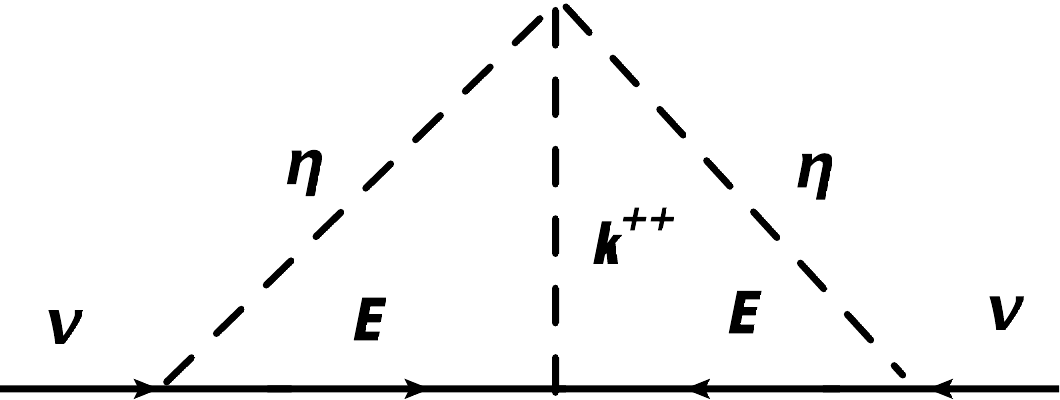}
\caption{One-loop diagrams generating neutrino mass.}
\label{fig:diagram-nu}
\end{figure}

In our model, neutrino masses are generated via two-loop diagram shown in Fig.~\ref{fig:diagram-nu}.
Here we write the Yukawa interactions for neutrino mass generation in mass basis such that 
\begin{align}
L &\supset 
 f_{i a} \bar \nu_{L_ i} E_{R_a}(s_\alpha H^+_1 + c_\alpha H^+_2) + g_{ab} \bar E_a E^c_b \nn\\
 &+\frac12 k^{++} (c^2_\alpha H^-_1H^-_1+s^2_\alpha H^-_2H^-_2 -2 s_\alpha c_\alpha H^-_1H^-_2)
 +{\rm h.c.},
\end{align}
where we simplify $g\equiv g_L=g_R$.
We then obtain neutrino mass matrix by calculating the diagram as 
\begin{align}
(m_\nu)_{ij} &=\sum_{a, b=1}^3  f_{ia} R_{ab} f^T_{bj},\quad  R_{ab}\equiv  R_{ab}^{(I)} + R_{ab}^{(II)}, 
\\
 R_{ab}^{(I)} &=2 \frac{\mu_{kss} s_\alpha^2c_\alpha^2 g_{ab}}{(4\pi)^4}
 \int\frac{[dx]_3}{y-1}  \int [dx']_3
 \left(
 \ln\left[\frac{\Delta^{H_1H_1}_{ab}}{\Delta^{H_1H_2}_{ab}}\right]
+
 \ln\left[\frac{\Delta^{H_2H_2}_{ab}}{\Delta^{H_1H_2}_{ab}}\right]
 \right),\\
 R_{ab}^{(II)} &= - \frac{\mu_{kss} s_\alpha^2c_\alpha^2 M_{E_a} g_{ab} M_{E_b}}{(4\pi)^4}
 \int\frac{[dx]_3}{y(y-1)}  \int [dx']_3
 \left(
 \frac{\Delta^{H_1H_2}_{ab} - \Delta^{H_1H_1}_{ab}}
 {\Delta^{H_1H_1}_{ab} \Delta^{H_1H_2}_{ab}}
+
 \frac{\Delta^{H_1H_2}_{ab} - \Delta^{H_2H_2}_{ab}}
 {\Delta^{H_2H_2}_{ab} \Delta^{H_1H_2}_{ab}} \right),\\
\Delta_{ab}^{H_i H_j}&= -x' \frac{x M^2_{E_a}+y m^2_{k}+z m^2_{H_i}}{y(y-1)} +y' M^2_{E_b} + z' m^2_{H_j},
\end{align}
and the neutrino mass matrix is diagonalized by a unitary matrix $V_{MNS}$ as $D_\nu = V_{MNS}^T m_\nu V_{MNS}$.
Since $R$ is a symmetric matrix with three by three, Cholesky decomposition can be done as $R= T^T T$, where $T$ is an upper-right triangle matrix. 
$T$ is uniquely determined by $R$ except their signs, where we fix all the components of $T$ to be positive signs~\footnote{To see more concrete form of $T$, see ref.~\cite{Nomura:2016run} for example.}.
Then, the Yukawa coupling $f$ is rewritten in terms of the other parameters as follows~\cite{Casas:2001sr}:
\begin{align}
\label{Eq:Yukawa-result}
f&= V_{MNS}^* D_\nu^{1/2} {\cal O} (T^T)^{-1} ,
\end{align}
where ${\cal O}$ is three by three orthogonal matrix with an arbitrary complex parameters.
Then Yukawa couplings $f_{ia}$ can have sizable values 
and significantly affect lepton flavor physics.

\subsection{$\ell_i \to \ell_j \gamma$ {and muon/electron $g-2$}}
The relevant interaction to induce $\ell_i \to \ell_j \gamma$ LFV process is obtained from second term of Eq.~\eqref{Eq:yuk} as
\begin{equation}
\label{Eq:intLFV}
  f_{ia}  \bar L_{L_i}\eta  E_{R_a}  + h \bar E_L e_R \chi + h.c. \supset f_{i a} \bar \ell_{L_i} E_{R_a} \eta^0 + h \bar E_L e_R \chi + h.c. .
\end{equation}
Note that $\eta^0$-$\chi$ mixing effect would provide significant contribution even if the mixing angle $\theta_s$ is small.
The diagram for $\ell_i \to \ell_j \gamma$(lepton $g-2$) including such a mixing is enhanced by mass of extra charged lepton $M_E$ due to chiral flip inside a loop.
Thus we include the mixing effect for $\ell_i \to \ell_j \gamma$ and muon/electron $g-2$.
Considering one loop diagrams, we obtain the BRs such that
\begin{align}
&{\rm BR}(\ell_i\to\ell_j\gamma)\approx\frac{48\pi^3\alpha_{em}C_{ij}}{G_F^2 }
\left( \left| (a_L)_{ij} \right|^2 + \left| (a_R)_{ij} \right|^2 \right),
\end{align}
where $C_{21}=1$, $C_{31}=0.1784$, $C_{32}=0.1736$, $\alpha_{em}(m_Z)=1/128.9$, and $G_F$ is the Fermi constant $G_F=1.166\times10^{-5}$ GeV$^{-2}$, and we assume to be $m_\eta \simeq m_{H_2}$ in evading oblique parameters.
The amplitudes are given by
\begin{align}
& (a_L)_{ij} = \frac{-1}{2(4\pi)^2} \sum_{a} 
\Bigl[
 \frac{m_{\ell_j} }{m_{\ell_i}} f_{j a} f^*_{i a} F(m_\eta,M_{E_a}) + 
h^\dag_{j a} h^T_{i a} F(m_\chi,M_{E_a})  \nonumber \\
& \hspace{4cm} - \sin \theta_s \frac{M_{E_a}}{m_{\ell_i}} h^\dag_{j a} f^\dag_{a i}  (F'(m_\chi,M_{E_a} ) - F'(m_\eta,M_{E_a} ) )
\Bigr], \\
& (a_R)_{ij} = \frac{-1}{2(4\pi)^2} \sum_{a} 
\Bigl[
  f_{j a} f^*_{i a} F(m_\eta,M_{E_a}) + 
\frac{m_{\ell_j} }{m_{\ell_i}} h^\dag_{j a} h^T_{i a} F(m_\chi,M_{E_a})  \nonumber \\
& \hspace{4cm} - \sin \theta_s \frac{M_{E_a}}{m_{\ell_i}} f_{j a} h_{a i}  (F'(m_\chi,M_{E_a} ) - F'(m_\eta,M_{E_a} ) )
\Bigr], \\
& F(m_a,m_b)\approx\frac{2 m^6_a+3m^4_am^2_b-6m^2_am^4_b+m^6_b+12m^4_am^2_b\ln\left(\frac{m_b}{m_a}\right)}{12(m^2_a-m^2_b)^4}, \\
& F'(m_a,m_b)\approx\frac{3 m^4_a-4 m^2_a m^2_b + m^4_b+ 4 m^4_a \ln\left(\frac{m_a}{m_b}\right)}{(m^2_b-m^2_a)^3},
\end{align}
where we have taken $\cos \theta_s \simeq 1$, $m_{H_1} \simeq m_\chi$ and $m_{H_2} \simeq m_\eta$ assuming $\theta_s \ll 1$.
The current experimental upper bounds are given by~\cite{TheMEG:2016wtm, Aubert:2009ag,Renga:2018fpd}
\begin{align}
{\rm BR}(\mu\to e\gamma)\lesssim 4.2\times10^{-13},\quad 
{\rm BR}(\tau\to e\gamma)\lesssim 3.3\times10^{-8},\quad
{\rm BR}(\tau\to\mu\gamma)\lesssim 4.4\times10^{-8},
\label{eq:lfvs-cond}
\end{align}
where we impose these constraints in our numerical calculation.

In addition, we obtain contribution to muon $g-2$, $\Delta a_\mu$, through the same amplitude taking $\ell_i = \ell_j = \mu$ that approximately gives 
\begin{equation}
\Delta a_\mu \simeq -m_\mu^2
\left[
(a_L)_{22} + (a_R)_{22}
\right],
\end{equation}
where $m_\mu$ is the muon mass.
{There is a discrepancy between the experimental results and the SM predictions at 3.3$\sigma$ level , and its deviation is given by $\Delta a_\mu=(26.1\pm8.0)\times10^{-10}$~\cite{Hagiwara:2011af}.}
In our numerical analysis, we also estimate the value.

 Here, we consider the same contribution to explain the $\Delta a_e$ so that this process does not affect to the other LFVs. This anomaly is recently reported by an experiment that suggests $\Delta a_e=-(8.8\pm 3.6)\times 10^{-13}$~{\cite{PYZEM}}. The point is opposite sign to the muon anomalous magnetic moment.  Then, this contribution is given by
\begin{align}
\Delta a_{e} = -m_e^2
\left[ (a_L)_{11} + (a_R)_{11} \right].
\end{align}  
Notice that the sign of $\Delta a_e$ can be different from that of $\Delta a_\mu$ due to scalar mixing term.

\subsection{Branching ratio of $\ell_i \to \ell_j \ell_k \bar \ell_l$}

\begin{figure}[tb]
\includegraphics[width=80mm]{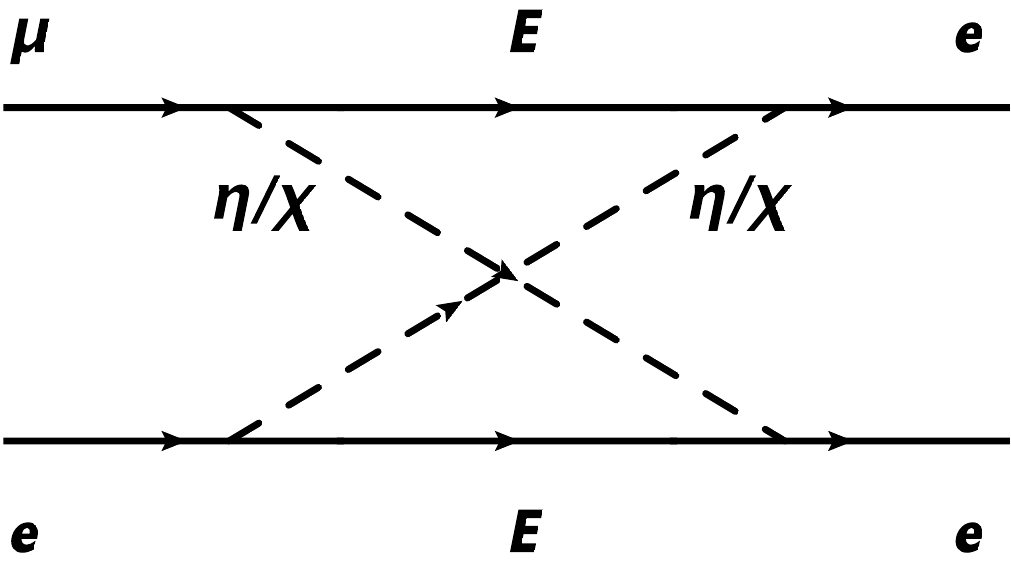}
\caption{The box diagram inducing $\ell_i \to \ell_j \ell_k \bar \ell_l$ decay and effective Lagrangian for $\mu e \to e e$ process.}
\label{fig:box-diagram}
\end{figure}

The LFV three body charged lepton decay processes are induced by box-diagram as shown in Fig.~\ref{fig:box-diagram}.
Calculating the one-loop diagram, we obtain BR for $\ell_i \to \ell_j \ell_k \bar \ell_l$ process such that 
\begin{align}
& {\rm BR}(\ell_i \to \ell_j \ell_k \bar \ell_l) \simeq \frac{m^5_{\ell_i} N_F}{6144 \pi^3 (4 \pi)^4 \Gamma_{\ell_i}} 
\left(\left|
f_{ja} f^\dagger_{ai} f_{kb} f^\dagger_{b l} G(m_\eta,M_{E_a},M_{E_b}) \right|^2
\right. \\ &\left.
+
\left|h^\dagger_{ja} h_{ai}   h^\dagger_{kb} h_{b l}G(m_\chi,M_{E_a},M_{E_b}) \right|^2
+\left|2
 f_{ka} M_{E_a} h_{ai} h^\dagger_{jb} M_{E_b} f^\dagger_{bl}H(m_\eta,m_\chi,M_{E_a},M_{E_b})\right|^2
 \right), \nn\\
& G(m_\eta,M_{E_a},M_{E_b}) = \int_0^1 \frac{\delta(x+y+z-1) x}{x m_\eta^2 + y M^2_{E_a} + z M^2_{E_b}} dx dy dz,\\
& H(m_\eta,m_\chi,M_{E_a},M_{E_b}) = \int_0^1 \frac{\delta(\alpha+\beta+\gamma+\delta-1) }
{(\alpha m_\chi^2 + \beta m_\chi^2 + \gamma M^2_{E_a} + \delta M^2_{E_b})^2} d\alpha d\beta d\gamma d\delta,
\end{align}
where $a,b$ are summed over $1-3$, $\Gamma_{\ell_i}$ is the total decay width of $\ell_i$, 
$N_F =2$ for $\ell_i \to \ell_j \ell_j \bar \ell_j$ or $\ell_i \to \ell_k \ell_k \bar \ell_j$ and $N_F =1$ for $\ell_i \to \ell_j \ell_k \bar \ell_k$~\cite{Crivellin:2013hpa}.
In this case we ignored $\eta^0$-$\chi$ mixing effect assuming $\theta_s \ll 1$, since we do not have enhancement factor in contrast to $\ell_i \to \ell_j \gamma$ case.
{In our numerical analysis, we impose current experimental constraints~\cite{Bellgardt1988,Hayasaka2010}:
\begin{align}
& BR(\mu^+\to e^+e^+e^-)\lesssim 1.0\times 10^{-12}, \quad BR(\tau^\mp\to e^\pm e^\mp e^\mp) \lesssim2.7\times 10^{-8}, \nonumber \\
& BR(\tau^\mp\to e^\pm e^\mp\mu^\mp) \lesssim1.8\times 10^{-8}, \quad BR(\tau^\mp\to e^\pm\mu^\mp\mu^\mp) \lesssim1.7\times 10^{-8} \nonumber \\
& BR(\tau^\mp\to \mu^\pm e^\mp e^\mp) \lesssim1.5\times 10^{-8}, \quad BR(\tau^\mp\to \mu^\pm e^\mp\mu^\mp) \lesssim2.7\times 10^{-8} \nonumber \\
& BR(\tau^\mp\to \mu^\pm\mu^\mp\mu^\mp) \lesssim2.1\times 10^{-8}.
\end{align}
}

\subsection{$\mu e \to e e$}
In our model $\mu e \to e e $ process in a muonic atom \cite{Koike2010} is also induced by Eq.~\eqref{Eq:intLFV}.
We then obtain relevant effective interactions from the same diagram inducing $\mu \to e \gamma$ and the box-diagram shown in Fig.~\ref{fig:box-diagram} such that
\begin{align}
\mathcal{L}_{\rm eff} = & - \frac{4 G_F}{\sqrt{2}} m_\mu (A_R \bar e \sigma^{\alpha \beta} P_R \mu + A_L \bar e \sigma^{\alpha \beta} P_L \mu) F_{\alpha \beta}  
-  \frac{4 G_F}{\sqrt{2}} [ g_3 (\bar e \gamma^\alpha P_R \mu)( \bar e \gamma_\alpha P_R e) \nonumber \\ 
& + g_4 (\bar e \gamma^\alpha P_L \mu)( \bar e \gamma_\alpha P_L e) + g_5 (\bar e \gamma^\alpha P_R \mu)( \bar e \gamma_\alpha P_L e)
+ g_6 (\bar e \gamma^\alpha P_L \mu)( \bar e \gamma_\alpha P_R e)
]+ h.c. \, ,
\end{align}
where the coefficients in our model are derived as
\begin{align}
& A_R \simeq \frac{e}{16 \pi^2} \frac{\sqrt{2}}{4 G_F} \sum_{a} 
\left(f_{1 a} f^*_{2 a} F(m_\eta,M_{E_a} ) 
+
 \frac{m_e}{m_\mu}  h^\dag_{1 a} h^T_{2 a} F(m_\chi,M_{E_a}) 
\right), \\
& A_L \simeq \frac{e}{16 \pi^2} \frac{\sqrt{2}}{4 G_F}  \sum_{a}
\left(
\frac{m_e}{m_\mu} f_{1 a} f^*_{2 a} F(m_\eta,M_{E_a}) 
+
 h^\dag_{1 a} h^T_{2 a}  F(m_\chi,M_{E_a}) 
\right), \\
& g_3 = \frac{\sqrt{2}}{128 \pi^2 G_F} \sum_{a,b} {(h_{1a}^\dag h_{2 a}^T) (h^\dag_{1b} h^T_{1 b})} G(m_\chi, M_{E_a}, M_{E_b}) \\
& g_4 = \frac{\sqrt{2}}{128 \pi^2 G_F} \sum_{a,b} {(f_{1a} f^*_{2 a}) (f_{1b} f^*_{1 b})} G(m_\eta, M_{E_a}, M_{E_b})   \\
& g_5 = g_6 = \frac{\sqrt{2}}{128 \pi^2 G_F} \sum_{a,b} \left[ {(f_{1a}M_{E_a} h_{a2}) (h_{1b}^\dag M_{E_b} f^\dag_{b1})} +{(f_{1a}M_{E_a} h_{a1}) (h_{1b}^\dag M_{E_b} f^\dag_{b2})} \right]  \nn \\
& \hspace{4cm}  \times H(m_\eta,m_\chi, M_{E_a}, M_{E_b}) 
\end{align}
By fixing $A_{L,R}$ and $g_i$ ($i=3-6$) values, we can determine the width of $\mu e \to e e$.
The ratio of the width to the total decay width of muonic atom, denoted by $R_{\mu^-e^-\to e^-e^-}$, is given by
\begin{align}
R_{\mu^-e^-\to e^-e^-}=&\frac{\tilde{\tau}_\mu G_F^2}{\pi^3}\int_{m_e}^{m_\mu-B_\mu^{1s}-B_e^{1s}}dE_1\left|\bm{p}_1\right|\left|\bm{p}_2\right| \nonumber\\
&\times\sum_{\kappa_1,\kappa_2,J}\left(2J+1\right)\left(2j_{\kappa_1}+1\right)\left(2j_{\kappa_2}+1\right)\left|A_LW_L+A_RW_R+\sum_{i=3}^{6}g_iW_i\right|^2,
\end{align}
where $\tilde{\tau}_\mu$ indicates the lifetime of a muonic atom~\cite{Suzuki1987}.
Here $B_\ell^{1s}$ ($\ell=\mu,e$) is the binding energy of the initial lepton $\ell$ in a $1s$ state.
For simplicity of the calculation, we consider only bound electrons in $1s$ states because of the small contribution from other bound electrons.
$E_n$ ($n=1,2$) is the energy of $n$-th emitted electron, which satisfies the energy conservation $E_1+E_2=m_\mu+m_e-B_\mu^{1s}-B_e^{1s}$.
$J$ is the total angular momentum of the lepton system, and $\kappa_n$ ($n=1,2$) is a nonzero integer which designates both the total and orbital angular momentum of the $n$-th electron, $j_\kappa$ and $l_\kappa$.
The expressions of $W_i$s ($i=L,R,3-6$) are given in Refs.~\cite{Uesaka2016,Uesaka2018}.

When we use a nucleus with a large atomic number, we get a larger transition rate of $\mu e\to ee$ \cite{Koike2010,Uesaka2016,Uesaka2018}.
To obtain a sizable $R_{\mu^-e^-\to e^-e^-}$, we assume $^{208}$Pb as a target nucleus in this analysis.

\subsection{Dark Matter}
In this paper, we consider DM relic density is explained by Yukawa interaction $\bar E_L \ell_R \chi$ where 
relevant annihilation process is $\chi \chi^* \to \ell \bar \ell$.
The cross section to explain the relic density is given by
\begin{align}
\sigma v \approx \frac{1}{192\pi}\sum_{i,j=e,\mu,\tau}\left|\sum_{a=1,2,3}h^\dag_{ia} h_{aj}\frac{m_\chi}{m_\chi^2+ M_{E_a}^2}\right|^2 v^2_{rel},
\end{align}
where we assume the massless limit of $e,\mu,\tau$ and set $v_{rel}^2\approx0.3$.
This is p-wave dominant and this cross section should be within the range of [1.77552-1.96967]$\times10^{-9}$
GeV$^{-2}$ at 2$\sigma$ confidence level in order to satisfy the correct relic density; 
{we relax this constraint in our numerical analysis as $[1.0-3.0]\times 10^{-9}$ GeV$^{-2}$ as an approximation.}

\section{Numerical analysis}

In this section we perform numerical analysis to search for allowed values of free parameters satisfying neutrino data and LFV constraints, 
and show ratios for LFV processes as well as electron/muon $g-2$ estimated by the allowed parameter sets.

In our numerical analysis, we scan relevant free parameters in our model in the following ranges:
\begin{align}
& m_{\chi} \in [100, 1000] \ {\rm GeV}, \  m_{H_1} \in [m_\chi, 10^{4}] \ {\rm GeV}, \  m_{H_2} \in [m_{H_1}, 10^{4}] \ {\rm GeV}, \  m_{k} \in [m_\chi, 10^{4}] \ {\rm GeV}, \nonumber \\
& m_{E_1} \in [m_\chi, 10^{4}] \ {\rm GeV}, \ m_{E_2} \in [m_{E_1}, 10^{4}] \ {\rm GeV}, \ m_{E_3} \in [m_{E_2}, 10^{4}] \ {\rm GeV}, \ \mu_{kss} \in [1, 10^4] {\rm GeV} \nonumber \\
& \sin \alpha \in [0.01, 1/\sqrt{2}], \ \sin \theta \in [10^{-6}, 0.01], \  g_{ab} \in [10^{-3}, \sqrt{4 \pi}], \nonumber \\
& |h_{a2}| \in [10^{-3}, \sqrt{4\pi}], \ |h_{ak}| \in [10^{-6}, \sqrt{4\pi}],
\end{align}
where $a=1-3$ and $k=1,3$. 
Note that we take $h_{a2}$ tends to be larger than  $h_{ak}$ in order to get sizable $\Delta a_\mu$. 
In our numerical analysis we require $\Delta a_\mu > 10^{-12}$ and $\Delta a_e < 0$.
We then search for the allowed parameter sets which satisfy LFV constraints discussed above and neutrino data from recent global fit~\cite{Esteban:2018azc,Nufit} 
\begin{align}
& | \Delta m^2_{\rm atm}|=[2.436-2.618]\times 10^{-3}\ {\rm eV}^2,  \quad  \Delta m^2_{\rm sol}=[6.79-8.01]\times 10^{-5}\ {\rm eV}^2,\nonumber \\
&\sin^2\theta_{13}=[0.02044-0.02435],  \quad  \sin^2\theta_{23}=[0.433-0.609],\nonumber \\
& \sin^2\theta_{12}=[0.275-0.350],
\end{align}
where we consider normal ordering (NO) case and Dirac(Majorana) CP phases are taken to be $[0, 2 \pi]$.
In our analysis, Yukawa couplings $f_{i \alpha}$ are obtained as an output value estimated by Casas-Ibarra parametrization given in Eq.~\eqref{Eq:Yukawa-result}. 

 \begin{figure}[t]
\includegraphics[width=80mm]{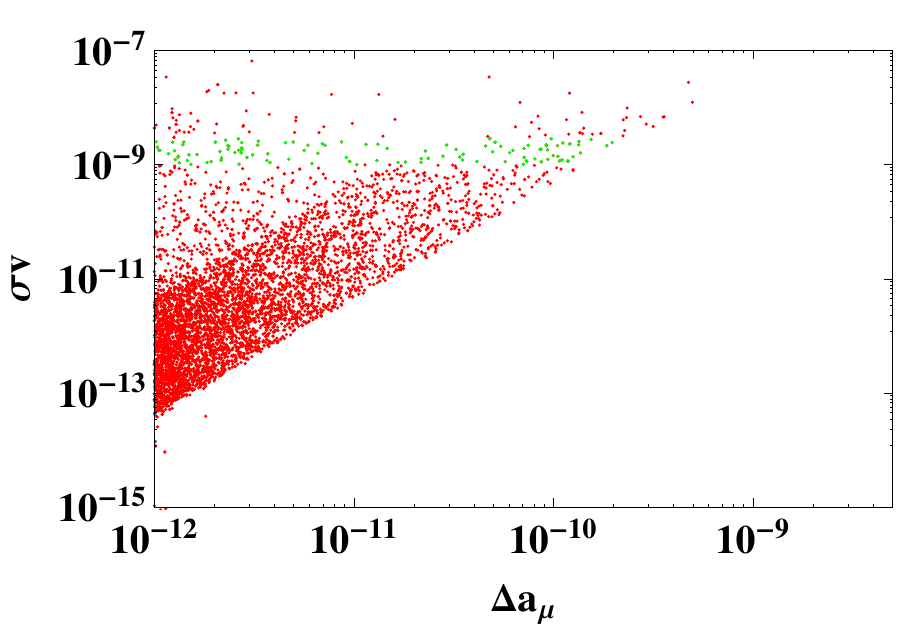}
\caption{Correlation between $\Delta a_\mu$ and $\sigma v$.}
\label{fig:amu-sigmav}
\end{figure}

 \begin{figure}[t]
\includegraphics[width=80mm]{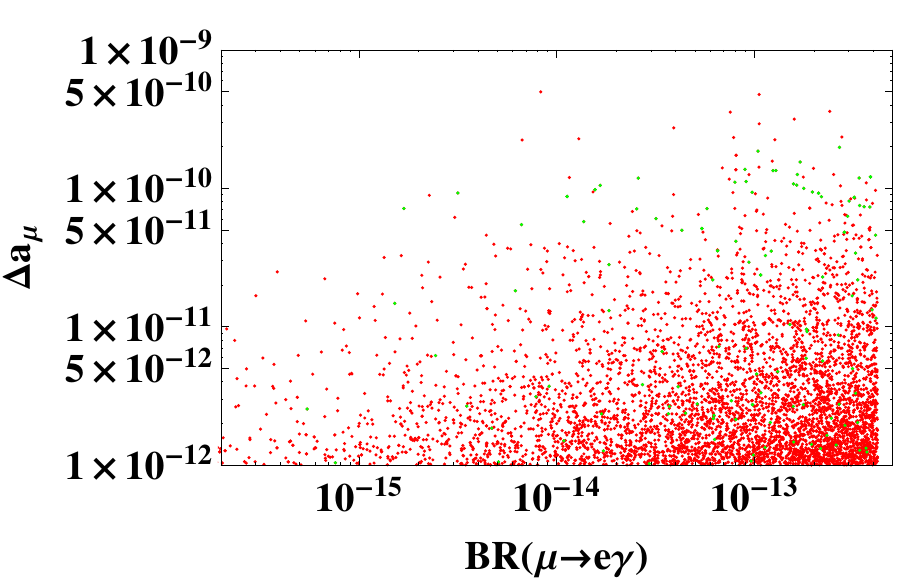} \
\includegraphics[width=80mm]{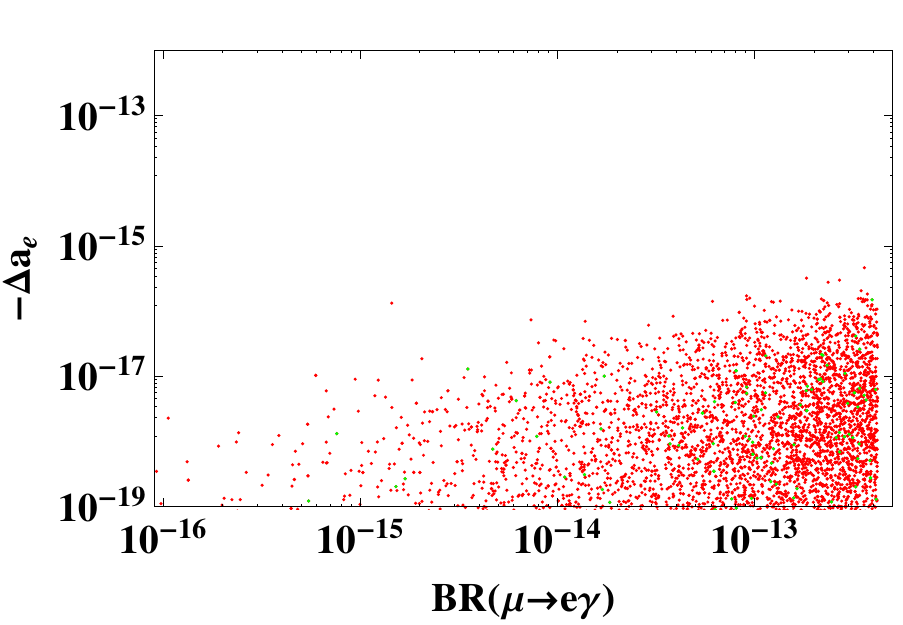}
\caption{Left: correlation for $\ell_i \to \ell_j \gamma$ and $a_\mu$ for allowed parameter sets. Right: correlation for $\ell_i \to \ell_j \gamma$ and $-a_e$ for allowed parameter sets.}
\label{fig:LFV1}
\end{figure}

 \begin{figure}[t]
\includegraphics[width=80mm]{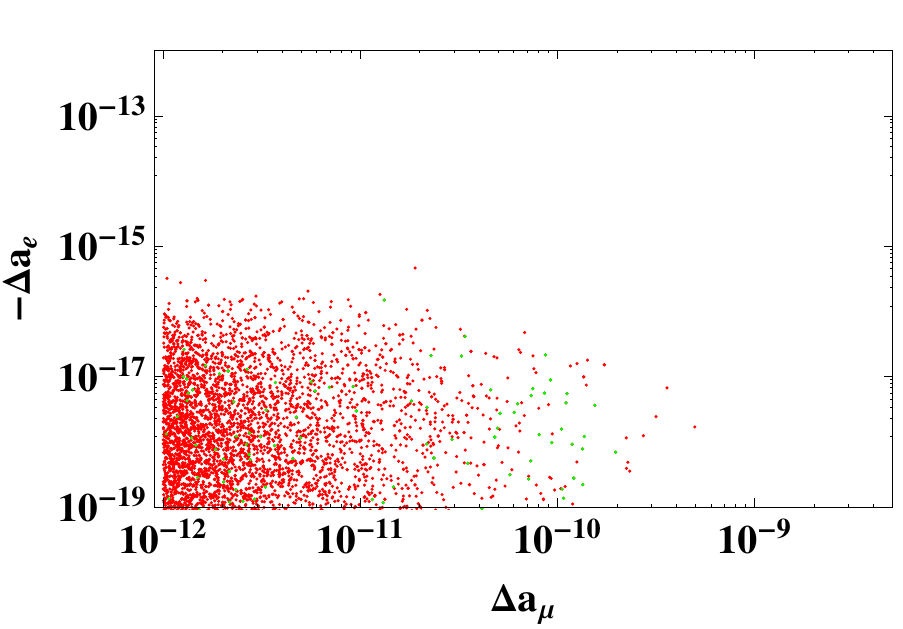}
\includegraphics[width=80mm]{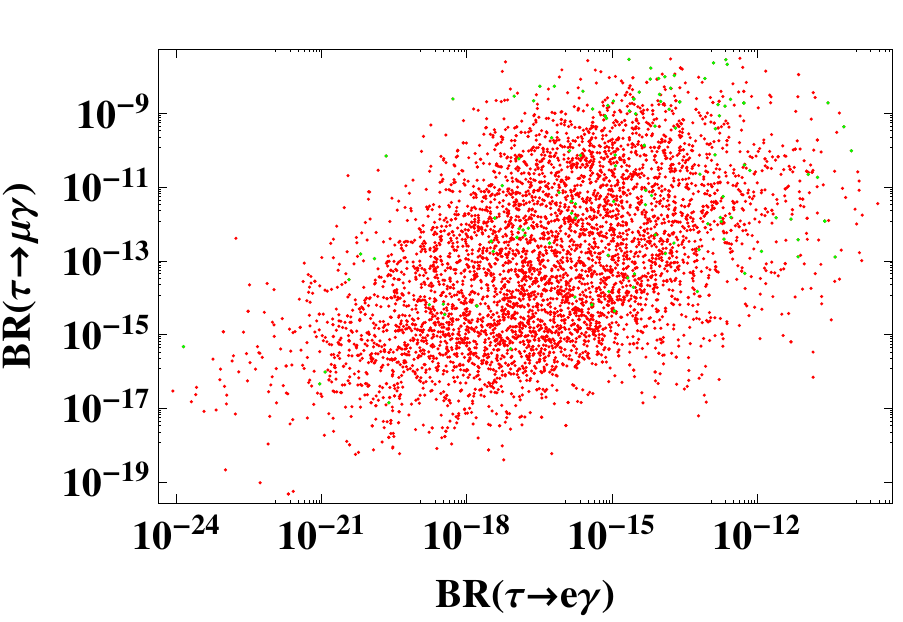}
\caption{Left: Correlation between muon and electron $g-2$. Right: Correlation between $BR(\tau \to e \gamma)$ and  $BR(\tau \to \mu \gamma)$}
\label{fig:LFV2}
\end{figure}

 \begin{figure}[t]
\includegraphics[width=80mm]{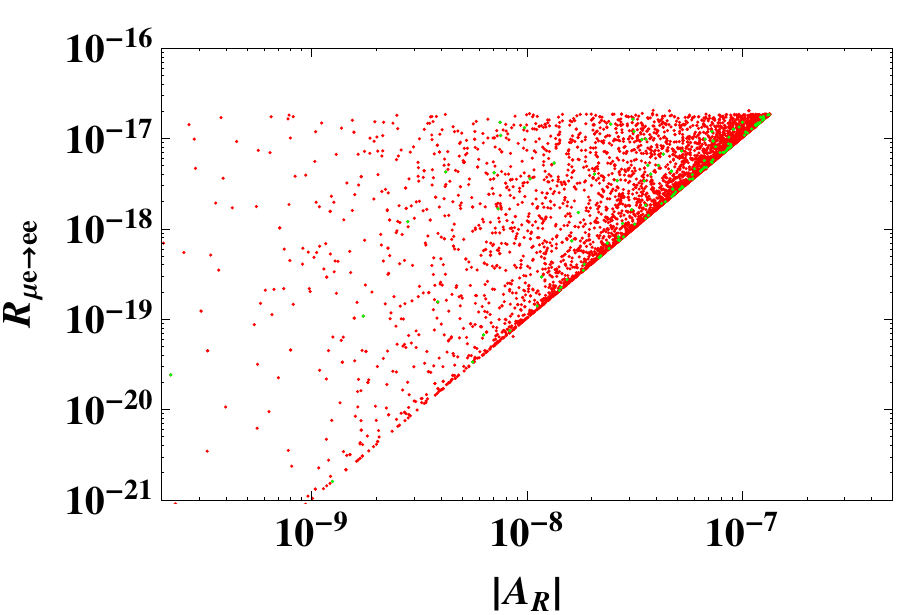} \
\includegraphics[width=80mm]{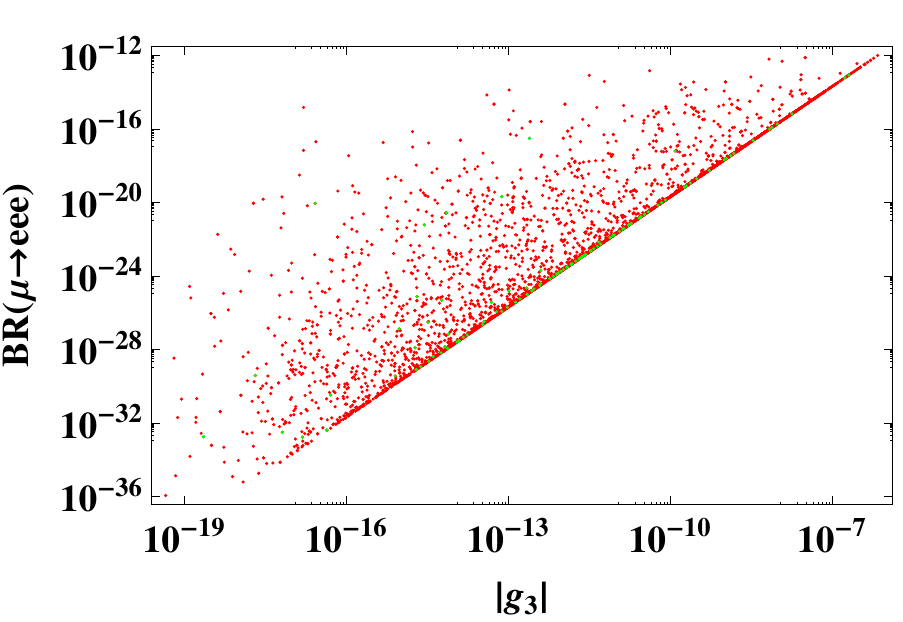}
\includegraphics[width=80mm]{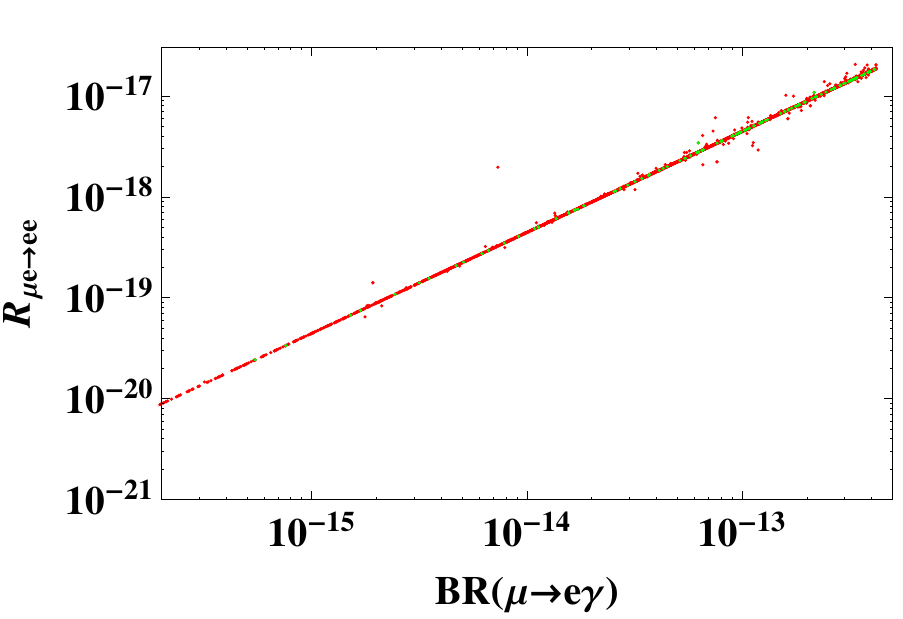} \
\includegraphics[width=80mm]{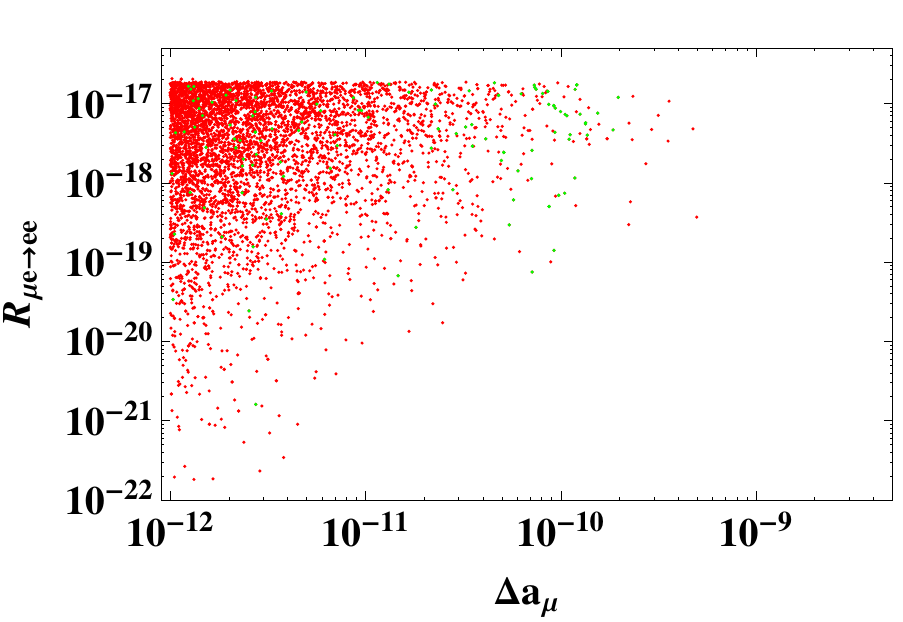}
\caption{Some correlations estimated with allowed parameter sets. Upper-left: correlation among $|A_R|$ and $BR(\mu \to e \gamma)$. Upper-right: correlation among $|g_3|$ and $BR(\mu \to eee)$. Lower-left: correlation among $BR(\mu \to e \gamma)$ and $R_{\mu e \to ee}$. Lower-right: correlation among $\Delta a_\mu$ and $R_{\mu e \to ee}$.}
\label{fig:LFV3}
\end{figure}

In the following, we show our observables estimated from parameter sets that are allowed by LFV constraints and neutrino data.
Fig.~\ref{fig:amu-sigmav} shows correlation between $\Delta a_\mu$ and $\langle \sigma v \rangle$.
We find that $\Delta a_\mu \sim 10^{-10}$ is preferred when annihilation cross section satisfies $10^{-9} \ {\rm GeV}^2 < \sigma v < 3.0 \times 10^{-9} \ {\rm GeV}^2$ giving observed relic density approximately; the region is indicated by green points and this presentation is used for following plots.
The muon $g-2$ can be up to $\sim 5 \times 10^{-10}$ when relic density of $\chi$ is smaller than observed one.
In left(right) plot of Fig.~\ref{fig:LFV1}, we provide estimated values of $BR(\mu \to e \gamma)$ and $\Delta a_\mu (- \Delta a_e)$  
showing correlation on $\{BR(\mu \to e \gamma), \Delta a_\mu (-\Delta a_e) \}$. 
We find that $BR(\mu \to e \gamma)$ tends to be larger for larger $|\Delta a_{e, \mu}|$. 
In fact, constraint from $\mu \to e \gamma$ restricts muon and electron $g-2$ and $|\Delta a_e|$ tends to be much smaller than observed value of $\sim 10^{-13}$.
In Fig.~\ref{fig:LFV2}, correlations on $\{\Delta a_\mu, -\Delta a_e \}$ and $\{BR(\tau \to e \gamma), BR(\tau \to \mu \gamma) \}$ plain are shown
in left- and right-panel.  
In Fig.~\ref{fig:LFV3}, we show some correlations among $R_{\mu e \to e e}$, Wilson coefficients $A_R$ and $g_3$, $BR(\mu \to e \gamma)$ and $\Delta a_\mu$. 
In most of the parameter sets,
{$R_{\mu e \to e e}$ is dominantly determined by the effect of $A_{L,R}$ indicated by upper-left plot, where upper limit of $|A_{L,R}|$ is determined by  
$BR(\mu \to e \gamma)$ constraint; $A_L$ and $A_R$ show similar behavior because the constraint requires upper limit of couplings $f$ and $h$ inducing them to be Max$[f]$ $\sim$ Max$[h]$.}
Thus $R_{\mu e \to e e}$ is also correlated with $BR(\mu \to e \gamma)$ as the lower-left panel.
The effect of $g_{3,4,56}$ is found as deviation from the correlation where upper limit of these Wilson coefficient is determined by the constraint of $BR(\mu \to eee)$ as shown in upper-right plot;
{behaviors of $g_{4}$ and $g_{56}$ are similar to $g_3$ because the constraint requires upper limit of couplings $f$ and $h$ inducing them to be Max$[f]$ $\sim$ Max$[h]$.}
We also find $R_{\mu e \to e e}$ tends to be large when $\Delta a_\mu$ is large {from the lower-right panel}.
The largest value of $R_{\mu e \to e e}$ is found to be $\sim 2 \times 10^{-17}$ which is obtained from maximal $A_{L}$ and $A_{R}$ values allowed by $BR(\mu \to e \gamma)$.
The expected number of stopped muons is estimated as $O\left(10^{17}\right)$ to $O\left(10^{18}\right)$ at the experiments for $\mu^--e^-$ conversion in near future, such as Mu2e \cite{Bartoszek2015} and COMET phase-II \cite{COMET2018}.
Thus we could obtain several number of events in these experiments, but they are planning to use an aluminum target, which is less suitable for $\mu^-e^-\to e^-e^-$ due to its small proton number.
In order to test the value of $R_{\mu e \to ee}$ with sufficient statistics,
we would need next generation experiments providing larger statistics or replacement of target materials to heavier nuclei.

\section{Summary}

We have investigated a model based on hidden $U(1)_X$ gauge symmetry in which neutrino mass is induced at two-loop level through interactions among particles in hidden sector and the SM leptons.
Generated neutrino masses are suppressed by two-loop factor and Yukawa couplings used in loop diagram can be sizable.
In addition, a scalar DM candidate is introduced that is stabilized by $Z_2$ symmetry as a remnant of $U(1)_X$ gauge symmetry.
Then we have formulated neutrino mass matrix, LFV processes, electron/muon $g-2$ and DM annihilation cross section which are induced via interactions among SM leptons and particles in $U(1)_X$ hidden sector.

We have carried out numerical analysis and searched for allowed parameter sets imposing neutrino data and current LFV constraints. 
Then we have discussed expected ratios for $\ell_i \to \ell_j \gamma$ , $\ell_i \to \ell_j \ell_k \bar \ell_l$ and $\mu e \to e e$, and electron/muon $g-2$ using allowed parameter sets.
In addition, we have estimated DM annihilation cross section which is given by interactions among DM, extra charged leptons and SM leptons.
We have found that the size of muon $g-2$ is preferred to be $\sim 10^{-10}$ when observed relic density can be obtained.
Furthermore LFV ratios tend to be large when muon $g-2$ is more than $10^{-10}$ and it could be tested in next generation experiments.

\section*{Acknowledgments}
The work is supported in part by KIAS Individual Grants, Grant No. PG054702 (TN) at Korea Institute for Advanced Study.
This research is supported by an appointment to the JRG Program at the APCTP through the Science and Technology Promotion Fund and Lottery Fund of the Korean Government. This was also supported by the Korean Local Governments - Gyeongsangbuk-do Province and Pohang City (H.O.). 
H. O. is sincerely grateful for the KIAS member. The work is supported in part by JSPS KAKENHI Grant Number JP18H01210 (Y.U.)

\appendix



\end{document}